\documentclass{article}

\usepackage{graphicx}%
\usepackage{multirow}%
\usepackage{amsmath,amssymb,amsfonts}%
\usepackage{amsthm}%
\usepackage{mathrsfs}%
\usepackage[title]{appendix}%
\usepackage{xcolor}%
\usepackage{textcomp}%
\usepackage{manyfoot}%
\usepackage{booktabs}%
\usepackage{algorithm}%
\usepackage{algorithmicx}%
\usepackage{algpseudocode}%
\usepackage{listings}%
\usepackage{bm}
\renewcommand*{\vec}[1]{\bm{#1}}
\usepackage{chemformula}
\usepackage{siunitx}
\usepackage{csquotes}
\usepackage{comment}
\usepackage{lineno}

\usepackage{lipsum}

\title{Altermagnetic magnon transport in the \textit{d}-wave altermagnet \ch{LuFeO3}} 
\author{Edgar Galindez-Ruales\\
Institute of Physics,\\ Johannes Gutenberg University Mainz, \\Staudingerweg 7, Mainz, 55128, Germany.\\
\and
Wanting Yang\\
Materials Genome Institute,\\ Institute of Quantum Science and Technology,\\ International Center for Quantum and Molecular Structures,\\ Shanghai University, \\99 Shangda Road, Shanghai, 200444, China.\\
\and
Tobias Dannegger\\
Fachbereich Physik, \\Universität Konstanz,\\ Universitätsstr. 10, Konstanz, D-78457, Germany.\\
\and
Moumita Kundu\\
Fachbereich Physik,\\ Universität Konstanz,\\ Universitätsstr. 10, Konstanz, D-78457, Germany.\\
\and
Jonas Köhler\\
Institute of Physics,\\ Johannes Gutenberg University Mainz,\\ Staudingerweg 7, Mainz, 55128, Germany.
\\
\and
Christin Schmitt\\
Institute of Physics,\\ Johannes Gutenberg University Mainz, \\Staudingerweg 7, Mainz, 55128, Germany.
\\
\and
Felix Fuhrmann\\
Institute of Physics, \\Johannes Gutenberg University Mainz, \\Staudingerweg 7, Mainz, 55128, Germany.
\\
\and
Akashdeep Akashdeep\\
Institute of Physics, \\Johannes Gutenberg University Mainz,\\ Staudingerweg 7, Mainz, 55128, Germany.
\\
\and
Duc Minh Tran\\
Institute of Physics,\\ Johannes Gutenberg University Mainz,\\ Staudingerweg 7, Mainz, 55128, Germany.
\\
\and
Xiaoxuan Ma\\
Materials Genome Institute, \\Institute of Quantum Science and Technology, \\International Center for Quantum and Molecular Structures,\\ Shanghai University, \\99 Shangda Road, Shanghai, 200444, China.\\
\and
Gerhard Jakob\\
Institute of Physics, \\Johannes Gutenberg University Mainz, \\Staudingerweg 7, Mainz, 55128, Germany.
\\
\and
Shixun Cao\\
Materials Genome Institute,\\ Institute of Quantum Science and Technology, \\International Center for Quantum and Molecular Structures, \\Shanghai University,\\ 99 Shangda Road, Shanghai, 200444, China.\\
\texttt{sxcao@shu.edu.cn}\\
\and
Ulrich Nowak\\
Fachbereich Physik,\\ Universität Konstanz,\\ Universitätsstr. 10, Konstanz, D-78457, Germany.\\
\texttt{ulrich.nowak@uni-konstanz.de}\\
\\
\and
Mathias Kläui\\
Institute of Physics,\\ Johannes Gutenberg University Mainz,\\ Staudingerweg 7, Mainz, 55128, Germany.\\
Center for Quantum Spintronics,\\ Norwegian University of Science and Technology,\\ Høgskoleringen 5, Trondheim, 7034, Norway.\\
\texttt{klaeui@uni-mainz.de}
}
\begin{document}
\maketitle

\begin{abstract}
Altermagnets exhibit a spin-split band structure despite having zero net magnetization, leading to special magnonic properties such as anisotropic magnon lifetimes and field-free spin transport. Here, we present a direct experimental demonstration of non-local magnon transport in the \textit{d}-wave altermagnet \ch{LuFeO3}, using both spin Seebeck and spin Hall effect-based injection and detection. We observe a non-local spin signal at zero magnetic field when the transport is along an altermagnetic direction, but not for transport along other directions. The observed sign reversal between two distinct altermagnetic directions in the spin Seebeck response demonstrates the altermagnetic nature of the magnon transport. In contrast, when transport is aligned along or perpendicular to the easy axis, both the first-harmonic signal and the sign-reversal effect vanish, consistent with symmetry-imposed suppression. These findings are supported by atomistic spin dynamics simulations, as well as linear spin wave theory calculations, which explain how our altermagnetic system hosts anisotropic spin Seebeck transport. Our results provide direct evidence of direction-dependent magnon splitting in altermagnets and highlight their potential for field-free magnonic spin transport, offering a promising pathway for low-power spintronic applications. 
\end{abstract} 






\maketitle

\section{Main}\label{sec1}

Magnon transport in magnetic materials has been shown to be a promising alternative to charge-based transport for low-dissipation spintronic applications \cite{chumak_magnon_2015, flebus_2024_2024}. The ability to propagate and manipulate spin information without net charge transport minimizes Joule heating, making magnon-based devices highly energy-efficient \cite{chumak_magnon_2015, chumak_advances_2022}. However, conventional magnon transport typically requires an external magnetic field to break degeneracies in the magnon dispersion, complicating device integration and scalability \cite{cornelissen_magnon_2016}. Recent theoretical and experimental advances have demonstrated that altermagnets—a newly identified class of collinear magnets—naturally exhibit spin-split magnon bands even in the absence of a net magnetization or applied magnetic field \cite{smejkal_chiral_2023}. This intrinsic property positions altermagnets as prime candidates for zero-field magnon transport, a key requirement for practical spintronic applications. Most known altermagnets are insulating materials, making magnon band splitting particularly relevant for magnonic spin transport applications \cite{bai_altermagnetism_2024}.

Altermagnets differ fundamentally from conventional antiferromagnets in that they host non-degenerate spin-split magnon bands due to their real-space symmetries \cite{smejkal_chiral_2023, liu_chiral_2024}.
This leads to alternating chirality splitting, which, combined with the approximately linear dispersion near the Brillouin zone center typical for antiferromagnets, significantly alters magnon dynamics and transport properties \cite{WeissenhoferMarmodoro2024, rezende_introduction_2019}.
Unlike other collinear ordered magnetic materials, altermagnets naturally support field-free spin transport due to the persistent alternating spin splitting of their magnon bands \cite{cui_efficient_2023}.
This intrinsic property has been predicted to enable spin Seebeck and Nernst effects, enabling efficient spin current generation without requiring magnetization control via external fields \cite{hoyer_spontaneous_2025}.

A distinctive feature of altermagnetic magnon transport is the strong directional anisotropy in magnon lifetimes and velocities resulting from the splitting \cite{costa_giant_2024}. Due to the momentum-dependent magnon band structure, specific crystallographic directions exhibit significantly different magnon decay rates, resulting in spatial anisotropy in magnon transport phenomena. This effect has been theoretically predicted to result in highly tunable spin transport properties, where certain transport directions can support efficient transport while other directions do not. The spin-split nature of these excitations opens avenues for controlled spin transport without requiring fields, magnetization gradients, or strong spin-orbit interactions. 

Because the \textit{g}-wave symmetry of conventionally used altermagnets, such as \ch{MnTe} \cite{jost_chiral_2025}, inherently prevents the generation of spin-polarized charge or magnon currents \cite{smejkal_chiral_2023}, one needs \textit{d}-wave altermagnets to generate spin currents under time-reversal symmetric conditions.
So far, only spin-polarized charge currents have been observed in conducting \textit{d}-wave altermagnets such as \ch{RuO2} \cite{bose_tilted_2022}.
Since most altermagnets are insulators, it is most important to establish altermagnetic magnon currents in \textit{d}-wave insulating altermagnets.
Rare-earth orthoferrites, such as \ch{YFeO3} and \ch{LuFeO3}, offer a combination of insulating behavior, symmetry conditions that enable \textit{d}-wave altermagnetism and low-damping magnon transport \cite{das_anisotropic_2022, bai_altermagnetism_2024}. These characteristics make them a highly versatile class of materials where both anisotropic spin wave propagation and long-range coupling effects can be engineered for future device applications, if altermagnetic magnon transport can be realized.


In this work, we present a direct experimental realization of non-local magnon transport in the insulating \textit{d}-wave altermagnet \ch{LuFeO3}. We use a spin Hall effect (SHE) and spin Seebeck (SSE) effect-based non-local geometry to inject and detect magnons. Our findings reveal a highly anisotropic thermal magnon transport response, with a sign reversal of the SSE signal for transport along distinct altermagnetic orientations. We demonstrate zero-field spin transport (SSE- and SHE-based), providing direct experimental confirmation that altermagnetic spin-split magnons can carry spin polarization without external fields or other external symmetry-breaking. These results establish altermagnets as a novel platform for controllable field-free, anisotropic magnon-based spin transport.

\subsection*{Altermagnetic properties of \ch{LuFeO3}}

\begin{figure}
\centering
\includegraphics[width=1.0\textwidth]{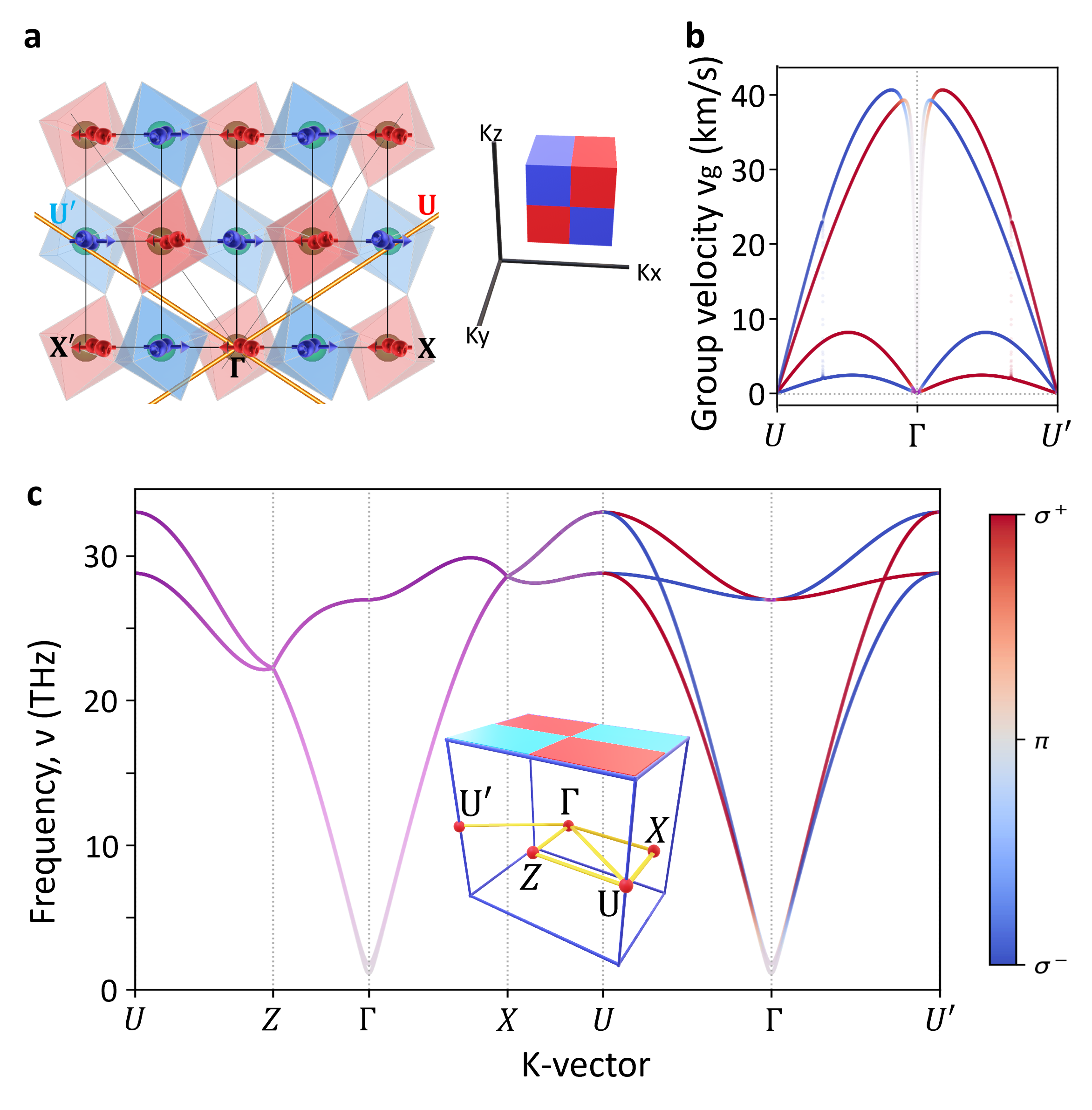}
    \caption{\textbf{Altermagnetic properties and magnon dispersion in \ch{LuFeO3}.}
\textbf{a.} Illustration of real- and reciprocal-space spin configurations in the orthoferrite \ch{LuFeO3}. The real-space viewing direction is a projection along the \textbf{b} axis and only shows the Fe atoms with their oxygen octahedra for clarity. The magnetic sublattices are antiparallel along the \textbf{a}-axis, with a small canting along the \textbf{c}-axis. The reciprocal-space structure reveals an alternating \textit{d}-wave symmetry in the \textbf{a}–\textbf{b} plane.
\textbf{b.} 
Calculated group velocity for the two non-degenerate magnon modes.
\textbf{c.} 
Calculated magnon dispersion along the altermagnetic and high-symmetry directions.
The color scale shows the magnon polarization (right-circular $\sigma^+$, left-circular $\sigma^-$, linear $\pi$).
Where both helicities occur, the branches are colored purple.
}\label{fig1:Theory}
\end{figure}

Figure \ref{fig1:Theory} illustrates the fundamental altermagnetic properties of \ch{LuFeO3}. In panel \ref{fig1:Theory}\textbf{a}, we depict the characteristic real-space and reciprocal-space spin-locking in the orthoferrite structure.
The magnetic order comprises four distinct sublattices oriented roughly antiparallel along the crystallographic \textbf{a}-axis, with additional small canting components along the \textbf{b}- and \textbf{c}-axes, the latter resulting in a weak net magnetization.
Notably, each sublattice exhibits a distinct local magnetization density caused by the altermagnetic symmetry of \ch{LuFeO3}.
This arrangement yields a characteristic alternating \textit{d}-wave symmetry in reciprocal space within the \textbf{a}-\textbf{c} plane.

The distortion in the local magnetization density results in a magnon spin-splitting along the $\Gamma$--$U$ directions, while such splitting is notably absent along the $\Gamma$--$X$ or $\Gamma$--$Z$ directions as shown in Fig. \ref{fig1:Theory}\textbf{c} (for details of the calculations, see Methods).
This direction-dependent magnon band splitting reflects the symmetry constraints of altermagnetic ordering, predicting distinct transport properties along specific crystallographic directions.

In the calculated magnon spectra (Fig. \ref{fig1:Theory}\textbf{c}) for a fixed $\vec{k}$-vector, there is a difference in the frequency (energy) value and the slope (group velocity) between the two different modes.
This, in the Bose-Einstein distribution, translates to differences in amplitude between the two magnon modes' occupation number, group velocities, decay length, and lifetime (see Extended Figure).
This difference is most pronounced along the identified altermagnetic directions $\Gamma$--$U$ and $\Gamma$--$U'$, and has an alternating nature.
At the Brillouin zone center ($\vec{k} = 0$), a relativistic splitting is observed, and the magnon modes exhibit elliptical polarization, transitioning gradually into more circular modes with increasing wave number. 
The group velocities of the two magnon branches (Fig. \ref{fig1:Theory}\textbf{b}) differ significantly even close to the center of the Brillouin zone. Such a difference in magnon group velocities and occupation motivates our non-local injection and detection approach, which relies on the excitation and propagation of magnons with characteristic anisotropic polarization, population, and velocity. 

\subsection*{Magnon transport via the SSE}

We first investigate the magnon transport in \ch{LuFeO3} by non-local injection via the SSE \cite{cornelissen_magnon_2016, lebrun_tunable_2018, lebrun_long-distance_2020, ross_propagation_2020, das_anisotropic_2022} and detection via the inverse spin Hall effect (i-SHE). In this method, a charge current flowing through a metal wire creates a temperature gradient, which generates magnons that propagate to the next Pt wire, where they are detected via the i-SHE. The i-SHE generates a current in the Pt wire if the excitation of the order parameter (Néel order or canted magnetization) has a perpendicular component to the charge current direction. Therefore, magnon detection is possible only when the detector wire has a component oriented perpendicular to the quantization axis \cite{lebrun_tunable_2018}.  Additionally, if the same number of right circular polarized magnons and left polarized magnons arrive, the total current generated via the i-SHE is zero. We refer to the magnon magnetization as the DC (time-averaged) spin-angular-momentum density carried by all magnon eigenmodes that contribute to transport. When a finite magnon magnetization reaches the detector wire, it translates into a perpendicular charge current in the Pt wire via the i-SHE, which is then measured.

\begin{figure}
\centering
\includegraphics[width=1.0\textwidth]{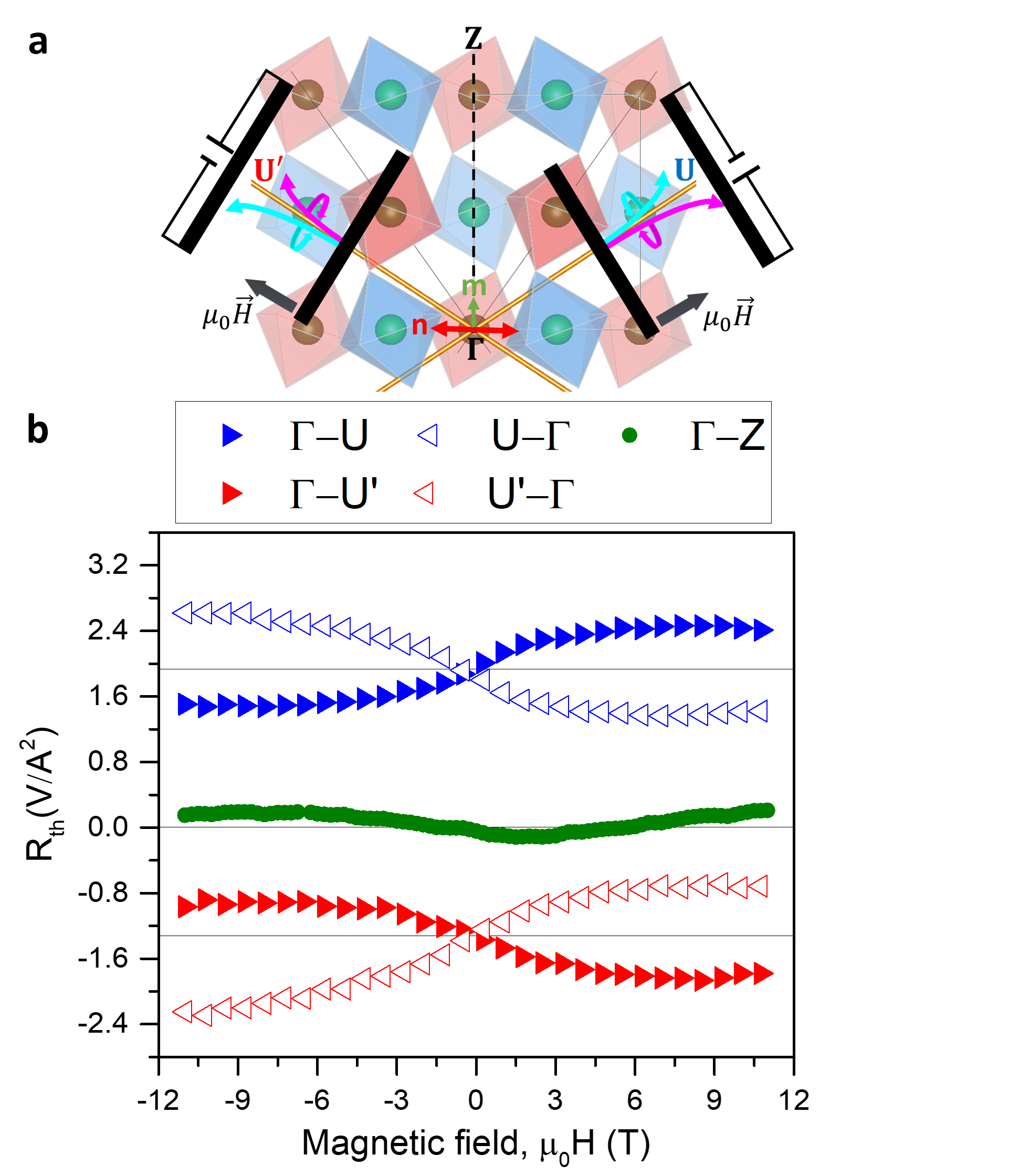}
\caption{\textbf{Altermagnetic non-local magnon transport via spin-Seebeck effect.}
\textbf{a.}
Schematic of the experiment, illustrating SSE signals along different (crystal-equivalent) transport directions. 
The relative amplitude and slope of right- and left-polarized magnons are represented schematically by larger and smaller arrows, respectively.
\textbf{b.} Non-local resistance signals as measured ($R_\mathrm{th}$) along two non-equivalent altermagnetic directions ($\Gamma$--$U$ and $\Gamma$--$U'$), compared to transport along the \textbf{c}-axis ($\Gamma$--$Z$).
The linear backgrounds of the signals have been subtracted. Error bars are smaller than the symbol size.
} \label{fig2:SSE}
\end{figure}

The SSE-driven spin transport is independent of the injected charge current polarity, as it originates from Joule heating, which generates thermal gradients rather than spin polarization. In our experimental setup, the heating from the injection wire generates a local temperature gradient that leads to the excitation of magnons up to $\nu\approx k_\mathrm{B} T$.  Due to the anisotropic nature of the altermagnetic magnon dispersion, a magnon magnetization is expected to occur only along certain crystal orientations determined by the reciprocal-space splitting.
The external magnetic field, the wires' orientations, and the resulting transport directions are chosen to be parallel to the altermagnetic direction in real space, i.e., normal to the real-space plane that corresponds to the $\Gamma$--$U$ reciprocal-space direction (Fig. \ref{fig2:SSE}\textbf{a}).

When comparing the measured magnon transport for the two different altermagnetic orientations in Figure \ref{fig2:SSE}\textbf{b}, we observe as a key feature a sign change in the non-local resistance when the transport is along $\Gamma$--$U$ and $\Gamma$--$U'$.
This means that, on average, magnons with opposite magnon magnetization reach the detector wire.
Reversing the direction of magnon transport (from $\Gamma$--$U$ to $U$--$\Gamma$) corresponds to reversing the propagation direction of magnons, and consequently, the detected SSE signal maintains the same overall sign.
Minor discrepancies observed between the signals in the forward and reverse directions (between $\Gamma$--$U$ and $U$--$\Gamma$) could stem from differences in spin mixing conductance at the detector interface or slight variations in injected thermal power.

Transport along high-symmetry directions (\textbf{a} and \textbf{c} axes) is used as a reference for our purposes. Along those directions, in orthoferrites (as previously probed in \ch{YFeO3} \cite{das_anisotropic_2022}), no sizable transport at zero magnetic field is observed, even along the crystallographic \textbf{c} direction ($\Gamma$--$Z$), which is oriented along the canted magnetic moment.
The magnitude of the non-local thermal signal along altermagnetic directions is one order of magnitude higher with respect to the signal along the \textbf{c}-axis.

The characteristic sigmoid curve of the observed field dependence can be explained by a gradual reorientation of the Néel vector up to the Néel-alignment field (around 3~T), with additional effects coming from the magnon ellipticity that changes with the applied field as long as the field exerts a torque on the Néel vector. The canted magnetization rotates with the field at fields below 0.3~T (Extended Figure). Above the Néel-alignment field, small changes might still occur due to an additional canting of the magnetic sublattices (Extended Figure) and ordinary magnetoresistance of the wires. This argument is in line with the spin Hall magnetoresistance (SMR) results (Extended Figure) from which a Néel-alignment field of $H_{\mathrm{c}}=3.1(3)$~T is extracted. In our previous work \cite{das_anisotropic_2022}, we showed that with an external magnetic field within the orthoferrite \textbf{a}-\textbf{c} plane, the Néel vector experiences a smooth reorientation until reaching perpendicular alignment rather than an abrupt transition; such a smooth transition is also observed in \ch{LuFeO3}. While the presence of a small canted magnetic moment in \ch{LuFeO3} could potentially contribute a small transport signal, our analysis reveals that this contribution is negligible compared to the dominant altermagnetic effects. This is particularly evident from high-field data and transport measurements conducted explicitly along the \textbf{c}-axis direction ($\Gamma$--$U$), which yield signals at least an order of magnitude lower than those along the altermagnetic directions. This contribution is observed at lower fields ($<0.22$ T), as present in our measurements (Extended Figure) and previously reported in the literature by J. Xu et al. \cite{Xu_2022_SSE-LuFO}. However, it is 20 times smaller than the observed signal along the altermagnetic directions.

\begin{figure}
\centering
\includegraphics[width=1.0\textwidth]{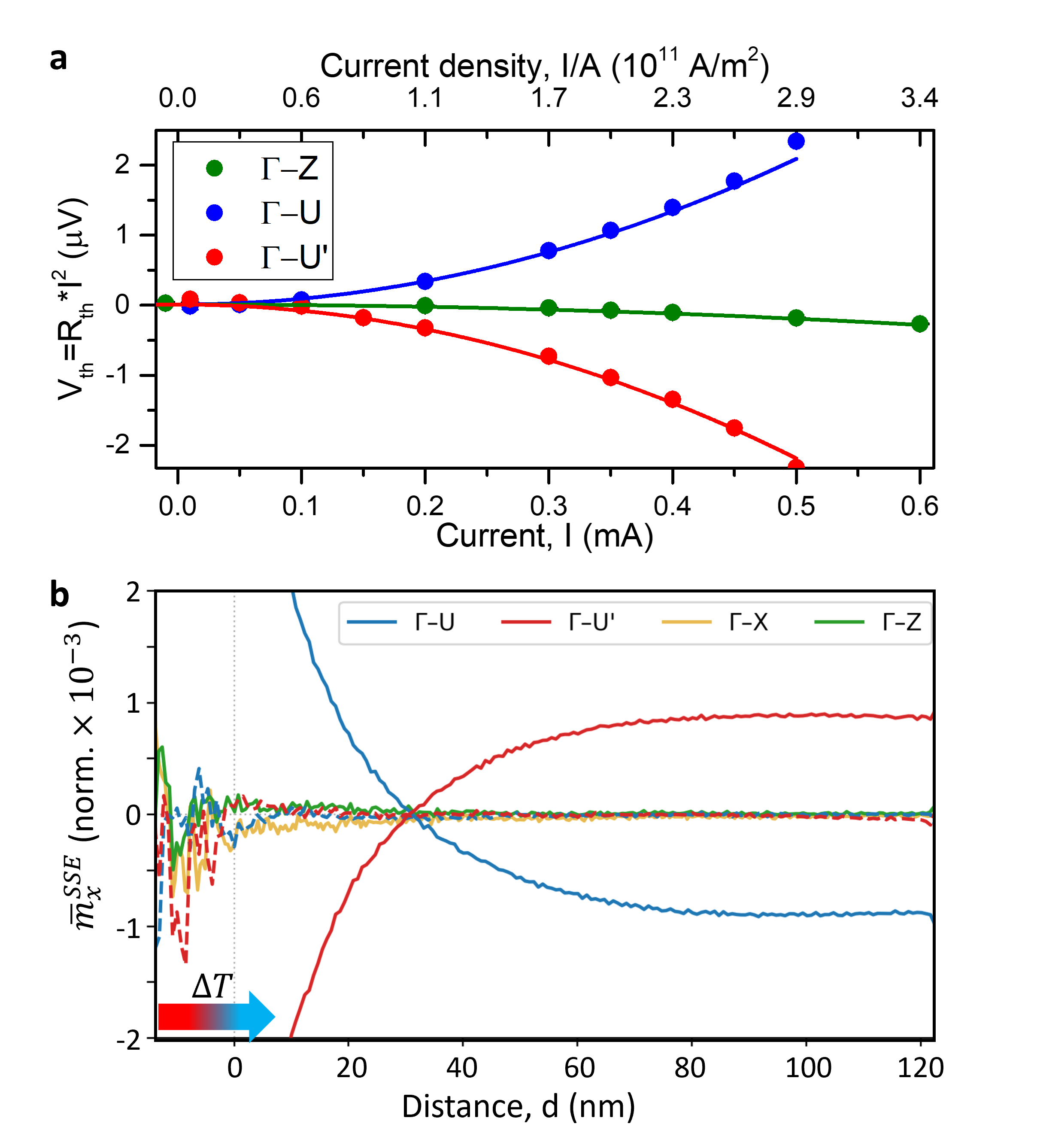}
\caption{\textbf{Spin Seebeck signal dependence.} \textbf{a.} The measured non-local SSE voltage as a function of the injected current at zero external magnetic field, with a corresponding quadratic fit. \textbf{b.} Atomistic spin dynamics simulation of the spin Seebeck effect along several directions. 
Shown is the $x$-component of the local magnetization resulting from a temperature step profile indicated by the red (300~K) and blue (0~K) colors.
The dashed lines correspond to the numerical results of the SSE transport with a non-altermagnetic version of the spin model (obtained by symmetrizing the exchange interactions along different directions).}
\label{fig3:Current_SSE-calculation}
\end{figure}

To check if the non-zero field observed signal is not a voltage artifact, we measure the non-local SSE voltage dependence on the injected electrical current, as shown in Figure \ref{fig3:Current_SSE-calculation}\textbf{a}. We find that the measured voltage scales quadratically with the current, indicating direct proportionality to the injected thermal power. Such a relationship confirms that the signal originates from thermally driven magnon transport, where Joule heating creates a temperature gradient, driving magnon currents via the SSE.

To understand the observed signal, we show in Figure \ref{fig3:Current_SSE-calculation}\textbf{b} SSE simulations \cite{Ritzmann2017_MagnonAccumulation} employing a temperature step function at the injection interface, designed to model experimentally induced temperature gradients.  In our atomistic spin simulations, a linear relationship was observed between the non-local SSE signal and the thermal gradient. Importantly, the simulations replicate the experimentally observed sign reversal in the SSE signal for the two altermagnetic crystal directions. This sign reversal directly reflects the intrinsic anisotropy in magnon dispersion and polarization unique to altermagnetic systems. In the direction of the crystal axes, $\Gamma$-$X$ and $\Gamma$--$Z$, the net transported spin is negligible in the model.

The dashed lines in Figure \ref{fig3:Current_SSE-calculation}\textbf{b} show the same transport simulations along altermagnetic directions but without the altermagnetic splitting in the isotropic interactions of the model, which results in vanishing net magnetization.
This clearly demonstrates that the observed spin-transport signal is exclusively due to altermagnetism.
Any additional relativistic components are negligible.

The magnon dispersion calculated from our spin model, shown in Figure \ref{fig1:Theory}, exhibits a chiral splitting of the magnon bands along the $\Gamma$-$U$ and $\Gamma$-$U'$ directions induced by the altermagnetic symmetry breaking (it is not present in the non-altermagnetic version of the model).
This frequency splitting, together with the corresponding group velocities and similarly anisotropic magnon lifetimes, leads in combination to anisotropic magnon decay lengths that are different for magnons of opposite helicity.
Together with differences in the thermal occupation of the magnon branches, this is what ultimately produces a finite spin accumulation in response to a temperature gradient along these directions.
The magnon modes that dominate far away from the heat source are the ones that are relatively close to, but not quite at, the $\Gamma$ point, where both thermal occupation, lifetime, and group velocity are high.

\subsection*{Magnon transport via the SHE}

In contrast to the SSE, spin-bias signals result directly from the spins injected by the SHE at the injector wire \cite{lebrun_tunable_2018}. This effect has a clear dependence on the current polarity. We differentiate between thermally driven magnon transport (SSE) and spin-bias (SHE) by analyzing polarity-independent (SSE) and polarity-dependent (SHE) signals. For spin-bias, the SHE in the Pt generates a spin current that accumulates at the Pt/insulator interface, creating a non-equilibrium distribution of magnons that propagates in the magnetic material until it reaches the detector wire, which absorbs them via the i-SHE. Here, the magnons are polarized, and the direction of the spins of the magnonic spin current with respect to the Pt wire determines whether a magnon can be injected or detected.

\begin{figure}
\centering
\includegraphics[width=1.0\textwidth]{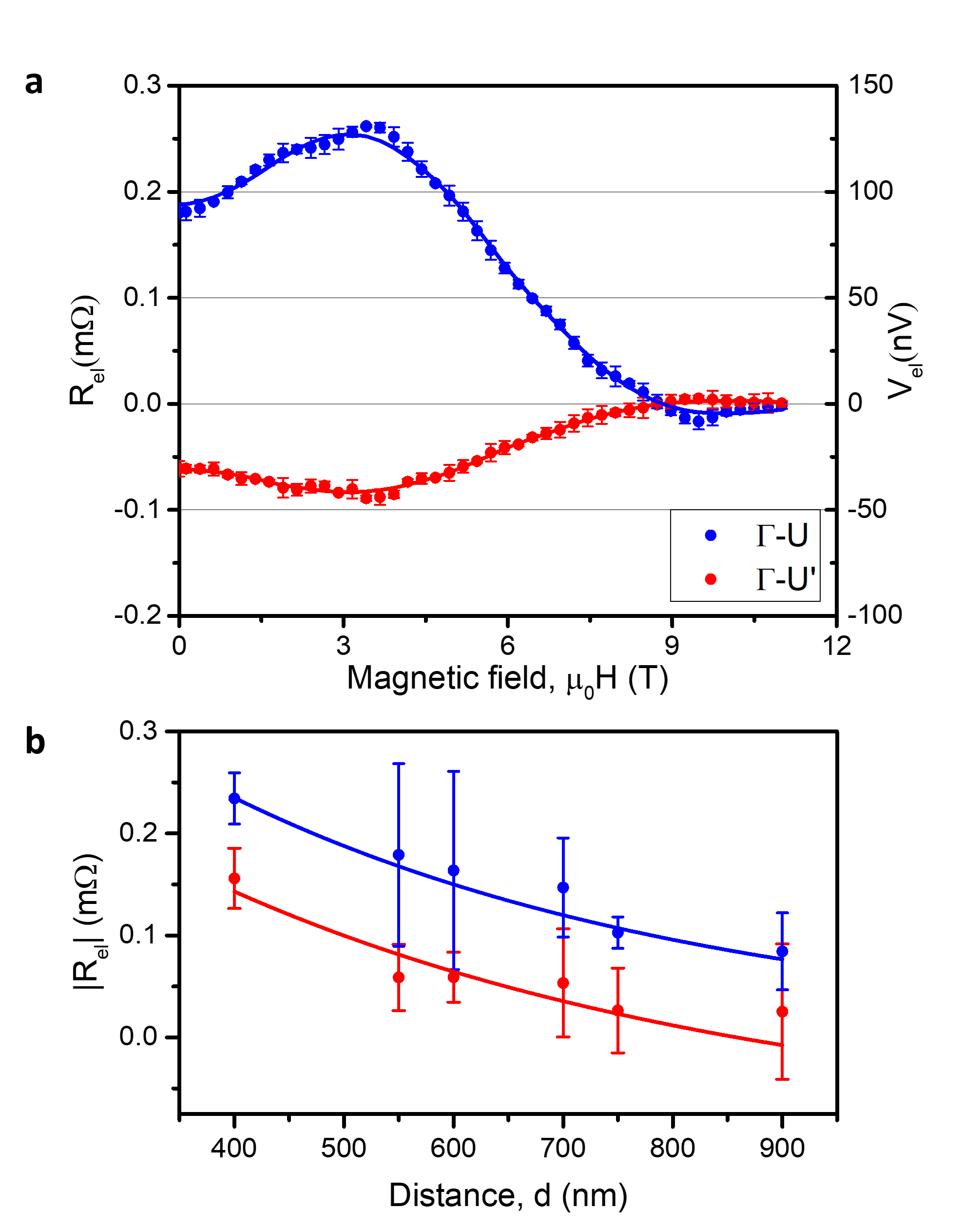}
\caption{\textbf{Spin transport in the altermagnetic directions.} \textbf{a.} Non-local spin transport along the two different non-equivalent altermagnetic directions. 
\textbf{b.} Non-local resistance as a function of distance between wires at zero external magnetic field, the detected voltage via the i-SSE follows an exponential decay with the distance between the wires.
}\label{fig4:1st_harmonic}
\end{figure}

Our experimental measurements reveal a distinct non-local spin-bias transport signal ($R_\mathrm{el}$) at zero external magnetic fields along the altermagnetic orientations (Fig. \ref{fig4:1st_harmonic}\textbf{a}). In contrast, no sizable signal at zero magnetic field is detected at highly symmetric orientations, where the altermagnetic contributions from $\Gamma$--$U$ and $\Gamma$--$U'$ cancel each other (\textbf{a} or \textbf{c} axis).
Finally, we quantify the propagation length; our measured non-local spin signal exhibits an evident exponential decay with increasing spatial separation between injector and detector contacts, yielding a spin diffusion length of approximately 600(59)~nm (Fig. \ref{fig4:1st_harmonic}\textbf{b}).
Furthermore, the spin bias signal scales linearly with the injected current, corroborating its direct relationship to spin accumulation generated by the SHE (Extended Figure), in contrast to the SSE signal that scales as $I^2$ (Fig. \ref{fig2:SSE}\textbf{a}).

Finally, for completeness, we want to point out that theory predicts different transport regimes and Néel vector orientations (Extended Figure). However, due to experimental limitations, we only probe one of these regimes here.
Both electrical ($R_{\mathrm{el}}$) and thermal ($R_{\mathrm{th}}$) transport signals measured in our experiments are proportional to the weighted sum of the DC (time-averaged) magnetization contributions from all excited magnon modes.

\section{Discussion}\label{sec3}

We have provided direct experimental demonstration of non-local altermagnetic magnon transport phenomena driven by intrinsic altermagnetism in \ch{LuFeO3}, highlighting distinct altermagnetic characteristics that conventional magnonic transport models cannot explain. A clear and robust sign inversion of the SSE signal between two different altermagnetic crystal directions, $\Gamma$--$U$ and $\Gamma$--$U'$, emerges as a central observed signature.
This sign reversal, arising from altermagnetic magnon band splitting, in conjunction with no such signal along other directions, confirms theoretical predictions and serves as a signature of \textit{d}-wave altermagnetism.

Our theoretical model shows that the SSE signal is caused by altermagnetic interactions and accurately predicts both the sign change of the SSE signal and its suppression when transport is measured along high-symmetry non-altermagnetic directions (easy or hard axes).
Experimental data confirm this suppression, underscoring that the observed spin transport signals along other directions predominantly originate from altermagnetic band-splitting, rather than from the bulk Dzyaloshinskii–Moriya interaction (DMI) in our sample.
Additionally, our experiments verify the expected proportionality between the measured non-local voltage and the temperature gradient, in excellent agreement with our theoretical framework.

Our findings establish the broad class of orthoferrites generalized as a playground for unconventional magnetism, enabled by their \textit{d}-wave magnetic symmetry, which facilitates controllable magnon-based spin transport. Unlike \textit{g}-wave altermagnets, such as the conventionally used MnTe, where magnon splitting emerges from sixfold rotational symmetry and is primarily characterized by chiral dispersion features, but by altermagnetic symmetry alone cannot lead to spin-polarized currents, the \textit{d}-wave symmetry in \ch{LuFeO3} offers significant practical advantages for spintronic applications. Specifically, the \textit{d}-wave symmetry yields strongly anisotropic magnon transport with zero-field functionality, exemplified by the robust sign reversal of the SSE signal between different altermagnetic directions. This directional anisotropy and intrinsic non-degeneracy in magnon bands enable field-free spin transport.
The combination of anisotropic and field-free magnon transport can be engineered for future device applications, establishing altermagnetic magnonics as a compelling new direction for future low-power spintronic technologies.

\section*{Acknowledgments}
This work has been supported by the German Research Foundation (DFG) under project No.\ 423441604.
T.D., Mo.Ku., and U.N. acknowledge additional support from the DFG through project No.\ 425217212 (SFB 1432, project B02).
All authors from Mainz acknowledge support from SPIN + X (DFG SFB TRR 173 No. 268565370, projects A01, A12, and B02). M.K. acknowledges support from the Research Council of Norway through its Centers of Excellence funding scheme, project number 262633 \enquote{QuSpin}.
M.K. and E.G.-R. also acknowledge the support and funding from the project CRC TRR 288 - 422213477 Elasto-Q-Mat (project A12).
This project has received funding from the European Union's Horizon Europe Programme, Horizon 1.2 under the Marie Skłodowska-Curie Actions (MSCA), Grant agreement No. 101119608 (TOPOCOM).
All authors thank Dr. Libor Šmejkal for providing insights into visualizing and determining the altermagnetic directions.

\section*{Declarations}
There is no conflict of interest or competing interests for any author.
\section*{Contributions}
M.K. conceived the idea and proposed the experiment.
T.D., Mo.Ku., and U.N. did the theoretical calculations and modeling.
W.Y. and X.M. fabricated the samples and performed the structural and magnetic characterization under S.C.'s supervision.
D.M.T. and A.A. deposited Pt on the single crystal for the device fabrication.
E.G.-R. fabricated the devices and performed the electrical measurements with the help of C.S., F.F., J.K., A.A., and W.Y.
E.G.-R analyzed the data and wrote the manuscript, with the help and discussion of C.S., F.F., J.K., T.D., U.N., G.J., and M.K.
S. C. acknowledges support from the National Natural Science Foundation of China (NSFC No. 12374116).
E.G.-R. and T.D. wrote the manuscript with input from W.Y., S.C., G.J., U.N., and M.K. All the authors reviewed the manuscript and commented on it.
\bibliographystyle{unsrt}
\bibliography{ArXiv}

\end{document}